\documentstyle[epsfig,12pt]{article}
\input amssymb.sty
\def\a{\alpha }   \def\b{\beta }
\def\dg{\dagger}
   
\def\t{\theta }                    
\def\d{\delta }     
\def\o{\omega }   
\def\l{\lambda }

\font\got=eufm10 scaled \magstep1
\font\gotscr=eufm7 scaled \magstep1
\font\gotscrscr=eufm5 scaled \magstep1
\newfam\gotfam
\textfont\gotfam=\got
\scriptfont\gotfam=\gotscr
\scriptscriptfont\gotfam=\gotscrscr
\def\got{\fam\gotfam}

\font\Bbb=msbm10 scaled \magstep1
\font\Bbbscr=msbm7 scaled \magstep1
\font\Bbbscrscr=msbm5 scaled \magstep1
\newfam\Bbbfam
\textfont\Bbbfam=\Bbb
\scriptfont\Bbbfam=\Bbbscr
\scriptscriptfont\Bbbfam=\Bbbscrscr
\def\Bbb{\Bbbfam}

\font\Cal=msbm10 scaled \magstep1
\font\Calscr=msbm7 scaled \magstep1
\font\Calscrscr=msbm5 scaled \magstep1
\newfam\Calfam
\textfont\Calfam=\Cal
\scriptfont\Calfam=\Calscr
\scriptscriptfont\Calfam=\Calscrscr
\def\Cal{\fam\Calfam}
\def\beq{\begin{equation}}
\def\eeq{\end{equation}}
\def\la{\langle}
\def\ra{\rangle}
   
\def\gappeq{\mathrel{\rlap {\raise.5ex\hbox{$>$}}
{\lower.5ex\hbox{$\sim$}}}}

\def\lappeq{\mathrel{\rlap{\raise.5ex\hbox{$<$}}
{\lower.5ex\hbox{$\sim$}}}}

\begin{document}

\begin{flushright}
{CERN-TH/99-177}
\end{flushright}
\vspace*{3mm}

\begin{center}

\Large{\bf TWO-PARTICLE CORRELATIONS FROM THE }

\vskip 0.1cm

\Large{\bf $q$-BOSON VIEWPOINT}

\end{center}

\vskip 0.5cm

\begin{center}
{\large {\sl D.V.~Anchishkin$^{\ a,\, b, \! \! }$
{\footnote{ E-mail: Dmitry.Anchishkin@cern.ch \ or/and \
                    Dmitry.Anchishkin@ap3.bitp.kiev.ua }}, \,
A.M.~Gavrilik$^{\ b, \! \! }$
{\footnote{ E-mail: omgavr@bitp.kiev.ua }}, \,
N.Z.~Iorgov$^{\ b, \! \! }$
{\footnote{ E-mail: mmtpitp@bitp.kiev.ua }
} }}

\vskip 0.6cm

$^a$CERN TH-Division, CH-1211 Geneva 23, Switzerland

\vskip 0.40cm

$^b$Bogolyubov Institute for Theoretical Physics \\
National Academy of Sciences of Ukraine \\
252143 Kiev-143, Ukraine\\

\end{center}

\vspace{3mm}

\begin{abstract}

We propose and develop to some extent a novel approach, which allows us to
effectively describe, for relativistic heavy-ion collisions,
the empirically observed deviation from unity of the intercept $\l$
(i.e. the measured value corresponding to zero relative momentum
${\bf p}$ of two registered identical pions or kaons) of the two-particle
correlation function $C(p,K)$.  The approach is based on the use
of two versions of the so-called $q$-deformed oscillators and
the corresponding picture of ideal gases of $q$-bosons.
By these techniques the intercept $\lambda$ is put into direct
correspondence with the deformation parameter $q$. For fixed
deformation strength, the model predicts dependence of the intercept
$\lambda$ on the pion pair mean momentum ${\bf K}$.

\end{abstract}

\vspace{3mm}


\section{Introduction}
\label{sec1}


The hadron matter under intense conditions of high temperatures and
densities has been extensively studied with the use of relativistic
heavy-ion collisions (RHIC).
The insight into the extreme matter with experimentally controlled
initial energies serves as an examination of existing models and theories.
On the other hand, RHIC promises to be a laboratory where a search for
new physics that extends beyond current imaginations can be made.

The models and approaches that are used to
describe the processes occurring in the reaction region are examined
by comparing provided predictions with experimental data on
single-, two- and many-particle momentum spectra, which contain
information on the source at the early stage (photons, dileptons)
and at the stage of so-called ``{\it freeze-out}'' (hadron spectra).
Two-particle correlations encapsulate information
about the space-time
structure and dynamics of the emitting source \cite{GKW}-\cite{heinz99}.
Usually, consideration of the correlations that occur in relativistic
heavy ion-collisions assumes that:
(i)  the particles are emitted independently (or the source is completely
     chaotic), and
(ii) finite multiplicity corrections can be neglected.
Then, correlations reflect
a) the effects from symmetrization
(antisymmetrization) of the amplitude to detect
identical particles with certain momenta, and
b) the effects that are generated by the final state interactions of the
detected particles between them and with the source.
At first sight, the final state interactions (FSI) can be regarded
as a contamination of ``pure'' particle correlations.
However, it should be noted that the FSI depend on
the structure of the emitting source and thus provide
information about source dynamics as well   \cite{anch98}.
Discussion of the latter is beyond the scope of the present paper.

The nominal quantity expressing the correlation function in terms
of experimental distributions \cite{boal} is
\begin{equation}
C({\bf k}_a,{\bf k}_b)=
\frac{\displaystyle P_2\left({\bf k}_a, {\bf k}_b\right) }
{\displaystyle P_1\left({\bf k}_a\right) \,
P_1\left({\bf k}_b\right) }
\ ,
\label{i1}
\end{equation}
\noindent where
$P_1\left({\bf k}\right) =E\, d^3N /d^3k$
and
$P_2\left({\bf k}_a, {\bf k}_b\right) = E_a \, E_b \ d^6N /(d^3k_ad^3k_b)$
are single- and two-particle cross-sections, ${\bf k}_a$ and ${\bf k}_b$
being on-shell asymptotic momenta.

In the absence of FSI, for a chaotic source, the correlation function
can be expressed as (see Appendix):
\begin{equation}
C(p,K) = 1\, + \,  \cos{\alpha} \,
\frac{
\left| \int  d^4 X \, e^{i  p\cdot X } S(X,K) \right| ^2 }
{
 \int  d^4 X \, S\Big( X,K+\frac{p}{2} \Big) \,
 \int  d^4 Y \, S\Big( Y,K-\frac{p}{2} \Big)
}
\ ,
\label{i2}
\end{equation}
where 4-momenta $K$ and $p$ defined as
\begin{equation}
K=\frac{1}{2} (k_a+k_b) \, , \ \ \ p=k_a-k_b \, .
\label{i3}
\end{equation}
The source function $S(x,K)$ (single-particle Wigner density) is defined
by emitted single-particle states $\psi_\gamma (t,{\bf x})$ at freeze-out
times:
 \begin{equation}
  S(Y,K) = \int d^4y\, e^{i K\cdot y}\,
  \sum_{\gamma , \gamma '} \rho_{\gamma \gamma '}\,
  \psi_\gamma \left(Y+{\textstyle{y\over 2}}\right) \,
  \psi_{\gamma '}^*\left(Y-{\textstyle{y\over 2}}\right) \, ,
 \label{i4}
 \end{equation}
where the summation (averaging) is taken over the set of all
quantum numbers $\{\gamma \}$
carried by the particle just before it is frozen out.
The source freeze-out density matrix $\rho_{\gamma \gamma '}$ is the
weight factor
of such an averaging and depends on the particular model of source,
for instance, thermal density operator is widely exploited.

For the system of identical particles that we are going to consider,
the two-particle wave function appears to be a symmetrized (antisymmetrized)
construction of  single-particle states (chaoticity assumption),
which reads
 \begin{equation}
   \psi_{\gamma_a \gamma_b}({\bf x}_a,{\bf x}_b,t_a) = \frac{1}{\sqrt{2}}
   \left[ \psi_{\gamma_a}({\bf x}_a,t_a)\,
          \psi_{\gamma_b}({\bf x}_b,t_a)
   + \, e^{i \alpha}
          \psi_{\gamma_a}({\bf x}_b,t_a)\,
          \psi_{\gamma_b} ({\bf x}_a,t_a) \right]
\, ,
 \label{i5}
 \end{equation}
where $\alpha =0$ for identical bosons, and $\alpha =\pi $ for identical
fermions.
Note that the function (\ref{i5}) is taken at freeze-out times (about
translation from detector times
$t \to \infty$ to emission times, see Appendix).

{}From now on we shall refer our consideration of two-particle
correlations    to identical bosons (pions, kaons, etc.).
As follows from Eq.~(\ref{i2}),
the boson correlation function should approach
the exact value two as the relative momentum approaches zero.
But as it was observed, from the very first
experimental data  and up to the most recent experiments,
the measured correlation function never reaches this value at ${\bf p}=0$.
To remove this discrepancy, the correlation function of identical bosons
is always taken in the form
 \begin{equation}
C(p,K) = 1\, + \,  \lambda \, f(p,K) \, ,
 \label{i6}
 \end{equation}
where $\lambda $ is drawn from an experimental fit to the data, usually in
the range $\lambda=$ 0.4\, --\, 0.9; $f(p,K)$ is commonly taken as a Gaussian
function (in any case, $f({\bf p}=0,K)=1$).
The deviation of $\lambda $ from unity in RHIC can be explained by
the production of secondary pions from
resonance decays which are outside the fireball.
The presence of long-lived resonances results in an increase of the measured
source size and life-times \cite{heiselberg96,heinz96}.

We are now coming to the  key idea of our paper.
Confining ourselves  to formula (\ref{i2})
to try to explain experimental data, it is then straightforward to put
into correspondence
the parameter $\lambda $  with the angle $\alpha $, so as to get,
by  means of
$\cos{\alpha}$, the right {\it reduction factor} $\lambda $.
Actually, in Eq.~(5) one can take the phase factor $e^{-i\a}$ in
place of the factor $e^{i\a}$. However, by simple algebra the
two-particle amplitude (52) can be reduced to the form
which results in the same two-particle probability as the former one.
Indeed, the correlation function (\ref{i2}), which is a measurable quantity,
is obviously symmetric with respect to $\alpha \to -\alpha$.
It turns out that an effective symmetrization of a two-particle
wave function in heavy-ion collisions exhibits similar features
to what one encounters in the description of the Aharonov-Bohm effect,
or in the physics of anyons.
This means that the two-particle wave function  of a boson pair
released from a dense and hot environment effectively acquires
an additional phase.
Hence, the drawn phenomenon can be ascribed to the properties of the
medium formed in RHIC, which, as we see, exhibits some non-standard
QFT behaviour through the considered correlation functions.
So,  adopting as a driving idea the fact that the correlation function
approaches $1+\lambda $ when the two-boson relative momentum approaches
zero, we will attempt to construct an effective model  capable
to mimic the real physical picture.
To perform this, we shall use as
our basic tool the so-called (algebra of) $q$-{\it deformed
commutation relations}, or techniques of {\it q-boson} statistics,
which certainly can be put in connection with the symmetrization rules.

The deformation parameter $q$ is viewed as an effective
(not universal) parameter which efficiently encapsulates most
essential features of complicated dynamics of the system under
study. In many cases, usage of appropriate $q$-algebra allows
one to reduce the treatment of complex system of interacting
particles to consideration of a system of non-interacting
ones at the price of complication (deformation) of the
commutation relations.
As an example let us mention the application
of $q$-deformed algebras to description of rotational spectra
of superdeformed nuclei \cite{Bo} -- here $q$ has different value
for each nucleus.

It is worth noting that in the context of hadron theory,
$q$-deformed algebras (or quantum algebras) were also already
applied. Such a usage
proved to yield a significantly improved description of hadron
characteristics, both regarding hadron scattering \cite{CA}-\cite{JKM}
-- nonlinearity of Regge trajectories -- and in the sector of such
static properties as hadron masses and mass sum rules
                                       \cite{Ga,GaIo}.

In what follows, we shall exploit, for the system of pions or kaons,
the (ideal) $q$-Bose gas picture based on two concrete versions
of $q$-bosons. From the viewpoint of direct physical meaning and/or
explanation of the true origin of the $q$-deformation in the considered
phenomenon, these versions differ from each other, first of all
in the question of whether $q$ must be real
or can also take such complex values as a pure phase.
Note that, at this stage, we do not go into details concerning
the diversity of other, than the above-mentioned,
``microscopical'' reasons (certainly, not completely unrelated)
for the appearance of $q$-deformed statistics.
Suffice it to mention that
the {\it composite nature of the particles} (pseudoscalar mesons)
under study may as well result                         \cite{GGG,AvKr}
in the $q$-deformed structures linked to the real deformation
parameter $q$.


\section{ The two versions of q-bosons }
\label{sec2}


In this section we give a brief sketch of main features
of the two versions (type ``A'' and type ``B'') of multimode $q$-oscillators,
which will be used in subsequent treatment.

    \vspace{4mm}

\hspace{68mm}
             \underline{\bf Type A}

    \vspace{4mm}
\noindent
The $q$-oscillators of this type
are defined by means of the relations \cite{ACKKD}
\smallskip
\[
[a_i,a_j]=[a^\dg_i,a^\dg_j]=0\ ,
\ \ \ [{\cal N}_i,a_j]=-\d_{ij} a_j\ ,\ \ \ \
[{\cal N}_i,a^\dg_j]=\d_{ij} a^\dg_j\ ,\ \ \ \
[{\cal N}_i, {\cal N}_j]=0\ ,
\]
\beq
a_i a_j^\dg-q^{\d_{ij}} a_j^\dg a_i=\d_{ij}\ .
\eeq
Note that, if $i\ne j$, this {\it system of independent}
$q$-oscillators differs
essentially from quons \cite{GM} whose different modes are non-commuting
(i.e. $q$-commuting).

{}From the vacuum state given by $a_i|0,0,\ldots\ra=0 \ $ for all $i$,
the state vectors
\beq
  |n_1,n_2,\ldots,n_i,\ldots\ra\equiv
\frac{1}{\sqrt{[n_1]![n_2]!\cdots[n_i]!\cdots}}
(a^\dg_1)^{n_1}(a^\dg_2)^{n_2}\cdots(a^\dg_i)^{n_i}\cdots|0,0,\ldots\ra
\eeq
are constructed as usual, so that
\beq
a^\dg_i|n_1,\ldots,n_i,\ldots\ra=\sqrt{[n_i+1]}|n_1,\ldots,n_i+1,\ldots\ra\ ,
\eeq
\beq
a_i|n_1,\ldots,n_i,\ldots\ra=\sqrt{[n_i]}|n_1,\ldots,n_i-1,\ldots\ra\ ,
\eeq
\beq
{\cal N}_i|n_1,\ldots,n_i,\ldots\ra=n_i|n_1,\ldots,n_i,\ldots\ra \ .
\eeq
Here the bracketed notation
\beq
[r]=\frac{1-q^r}{1-q}\ ,\ \ \ \
{\rm along \ with} \ \ \ \
[r]!=[1][2]\cdots[r-1][r]\ ,\ \ \ [0]!=1 ,
\eeq
is used. The $q$-bracket $[A]$ for an operator $A$ is understood
as a formal expression (formal series). At $q\to 1$, from $[r]$ and $[A]$
we recover $r$ and $A$, thus returning to the formulas for the standard
bosonic oscillator. In what follows it will be assumed that
\beq
-1\le q\le 1\ ;
\eeq
for each such value of the {\it deformation parameter} $q$,
the operators $a^\dg_i\ ,$ $a_i$ are mutual conjugates.

In the generic case where $q\ne 1$,  the bilinear $a^\dg_i a_i$
does not equal the number operator ${\cal N}_i$ (as this is true
for usual bosonic oscillators) but, instead,
\beq
a^\dg_i a_i=[{\cal N}_i] .
\eeq
The inverse of the latter relation is given by the formula \cite{CKSSS}:
\beq
{\cal N}_i=\sum^\infty_{s=1}\frac{(1-q)^s}{1-q^s} (a^\dg_i)^s a_i^s
\, ,
\eeq
expressing the number operator as a (formal) series of
creation and annihilation operators.

    \vspace{4mm}
\hspace{68mm}
            \underline{\bf Type B}

   \vspace{4mm}
\noindent
The $q$-oscillators of the second
type are defined through the relations \cite{BM, AFZMP}:
\[
[b_i,b_j]=[b^\dg_i,b^\dg_j]=0\ ,\ \ \ [N_i,b_j]=-\d_{ij} b_j\ ,\ \ \ \
[N_i,b^\dg_j]=\d_{ij} b^\dg_j\ ,\ \ \ \
[N_i, N_j]=0 \ ,
\]
\beq
b_i b_j^\dg-q^{\d_{ij}} b_j^\dg b_i=\d_{ij}q^{-N_j}\ ,\ \ \ \
b_i b_j^\dg-q^{-\d_{ij}} b_j^\dg b_i=\d_{ij}q^{N_j} \ .
\eeq
Again we have
\beq
b^\dg_i b_i=[N_i]
\eeq
where the notation for the $q$-bracket this time means:
\beq
[r]=\frac{q^r-q^{-r}}{q-q^{-1}} \ .
\eeq
Formulas completely analogous to Eqs. (8)--(11) are valid also for the
operators $b_i, \ b_j^\dg \ $ if, instead of (12), we now  use
the definition (18) for $q$-brackets.
Clearly, the equality $b^\dg_i b_i = N_i$ holds only in the
``no-deformation'' limit of $q=1$. For consistency of the conjugation,
it is required that either $q$ is real or
\beq
q=\exp (i \t)\ ,
          \ \ \ \ \ \ \ \ \ \ 0 \le \t < 2\pi \ .
\label{19}
\eeq

In what follows we will consider the type ``A'' as well as the type
``B'' oscillators.
For the type ``B'' oscillators the exponent form (\ref{19}) will be
adopted (compare with the symmetrization phase $\alpha $ in
Eq.~(\ref{i5})).


\section{ Statistical  $q$-distributions }
\label{sec3}


For the dynamical multi-pion or multi-kaon system, we consider
the model of ideal gas of $q$-bosons (IQBG)
taking the free (non-interacting) Hamiltonian in the form \cite{MMGT,AG}
\beq
H=\sum_i{\o_i {\cal N}_i}
\, ,
\eeq
where $\omega_i=\sqrt{m^2+{\bf k}_i^2}\ $, ${\cal N}_i$ is defined
as above, and subscript $i$ labels energy eigenvalues.
It should be emphasized that
among a large variety of possible choices of Hamiltonians, this is
the unique truly non-interacting one, which possesses
an additive spectrum. From now on, we assume that $3$-momenta of
particles take their values from a discrete set (i.e. the system
is contained in a large finite box of volume $\sim L^3$).

As usual, basic statistical properties are obtained by evaluating
thermal averages such as
\[
\la A \ra=\frac{{\rm Sp}(A\rho)}{{\rm Sp}(\rho)}\ ,\ \ \ \ \
\rho=e^{-\b H} \, ,
\]
where $\b=1/T$ and the Boltzmann constant is set equal to 1.
The averaging here is taken with respect to the chosen Hamiltonian (20).

It is an easy task to calculate the quantity $\la q^{{\cal N}_i} \ra $,
and to obtain
\beq
\la q^{{\cal N}_i} \ra =
\frac{e^{\b\o_i}-1}{e^{\b\o_i}-q}\ .
\eeq
{}From this we find the distribution function (recall that $q$ is
from the interval $\ -1\le q\le 1 $):
\beq
\la a_i^\dg a_i \ra=\frac{1}{e^{\b\o_i}-q}\ .
\eeq
In the no-deformation limit $q\to 1$, this reduces to the
Planck-Bose-Einstein distribution, as it should, since at $q=1$
we return to the standard system of bosonic commutation relations.

At $q=-1$ and $q=0$, the distributions we get coincide
respectively with Fermi-Dirac and Maxwell-Boltzmann ones. It should be
emphasized that this coincidence is rather formal: the defining
relations (7) at $q=-1$ or $q=0$ differ from those for the system
of fermions or the non-quantal (classical) system. The formal
coincidence of Eq.~(22) at $q=-1$ with a Fermi-Dirac distribution
can be interpreted \cite{Sa} as due to the impenetrability
(the hard-core property) of such bosons.
The difference with the system of genuine fermions lies
in {\it commuting} (versus fermionic anticommuting) of
{\it non-coinciding} modes at $q=-1$; see (7).

Let us now turn to the type ``B''  $q$-bosons. The Hamiltonian
is chosen again as that of IQBG with the number operator defined in
(16) and (17), i.e.,
\beq
H=\sum_i{\o_i  N_i}\ .
\eeq
Calculation of $\la q^{\pm N_i} \ra $ yields
\beq
\la q^{\pm N_i} \ra =
\frac{e^{\b\o_i}-1}{e^{\b\o_i}-q^{\pm 1}} \ .
\eeq
{}From the relation
\[
\la b_i^\dg b_i \ra=\frac{1}{e^{\b\o_i}-q}\la q^{-N_i} \ra
\]
or, equivalently,
$
\la b_i^\dg b_i \ra=\la a_i^\dg a_i \ra_{\rm BE} \la q^{N_i} \ra
\la  q^{-N_i} \ra ,
$
we obtain the formula for the $q$-deformed distribution function
(taking into account that $q+q^{-1}=[2]=2\cos\t$):
\beq
\la b_i^\dg b_i \ra=\frac{e^{\b\o_i}-1}
{e^{2\b\o_i}-2\cos(\theta)e^{\b\o_i}+1} \ .
\eeq
Note that, although the deformation parameter $q$ is chosen in a
particular {\it complex} form, see (19), the explicit expression
for the $q$-distribution function turns out to be real, owing to its
specific dependence on $q$ through the combination $q+q^{-1}$.

The shape of the function $f(k)\equiv \la b^\dg b \ra (k) $
from (25) corresponding to the gas of pions
modelled by IQBG is picture\footnote
{
The (isotriplet-averaged) pion mass $m(\pi^{\pm,0})=139.57$~MeV
and the temperature $T=120$~MeV are taken as inputs.
The deformation value, encoded in $\cos\t$, is fixed to be $\t=24^\circ$. }
in Fig.~1 (curve II). For comparison, the standard
Bose-Einstein distribution function (curve I) and the classical
Maxwell-Boltzmann one (curve III) are also presented in the same figure.
As is clearly seen, the $q$-deformed
distribution function lies completely
in between the other two curves, thus demonstrating\footnote
{
          The same is true also for the (apparently more simple)
          $q$-distribution (22) of the type ``A'' $q$-bosons. }
that the deviation of the $q$-distribution (25) from the quantum
Bose-Einstein distribution goes in the ``right direction'', towards the
classical Maxwell-Boltzmann one.

\vspace{-0.2cm}

\bigskip
\begin{figure}[ht]
\begin{center}
\epsfig{file=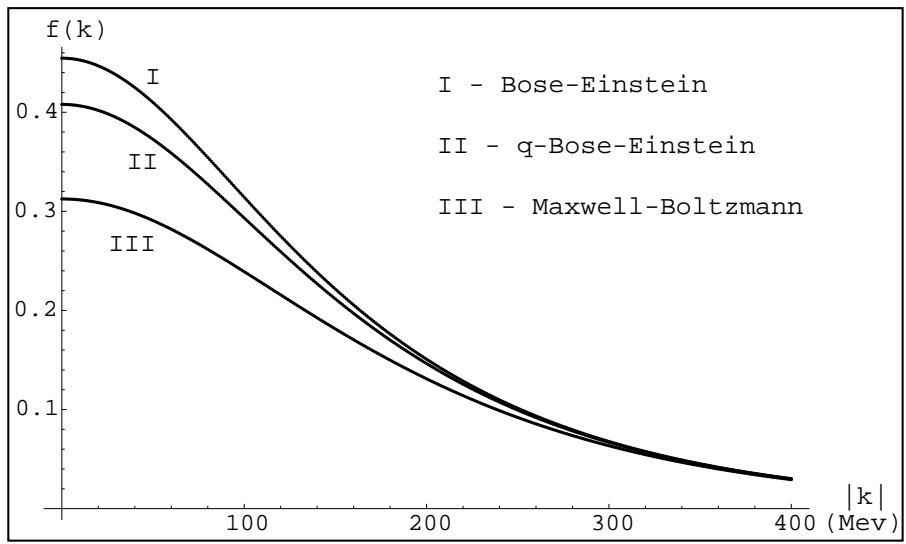,width= 12cm,angle=0}
\vspace{0.1cm}
\caption{ The $q$-distribution function (25) versus momentum
(curve II), in comparison with the quantum Bose-Einstein (curve I)
and classical Maxwell-Boltzmann (curve III) distributions. The inputs are:
$T=120$ MeV, $m=m_\pi\ $; curve II corresponds to the deformation angle
$\t=24^\circ$. }
\end{center}
\end{figure}

\vspace{-0.1cm}

Analogous curves for $q$-distribution functions, with similar
properties, can be given for other fixed data. Let us remark
that, in the case of kaons, because of their larger mass
and higher empirical value of the intercept $\l\simeq 0.88$
(which corresponds to the smaller deformation in our model),
such a curve should lie significantly
closer to that of the Bose-Einstein distribution.

  It is worth noting that the $q$-distribution functions (22) and (25)
already appeared (in the context of thermal field theory) in \cite{AG}.
We give them here for completeness of exposition, since they are
connected with new results to be described in the next section.


\section{ Two-particle correlations of $q$-bosons }
\label{sec4}


Let us now turn  to the
issue of two-particle correlations.
{}From the easily verifiable identity
\[
a_i^\dg a_j^\dg a_k a_l -  q^{-\d_{ik}-\d_{il}}  a_j^\dg a_k a_l a_i^\dg
=[a_i^\dg , a_j^\dg] a_k a_l + a_j^\dg [a_i^\dg , a_k]_{q^{-\d_{ik}}} a_l
+q^{-\d_{ik}}  a_j^\dg a_k [a_i^\dg,a_l]_{q^{-\d_{il}}} \ ,
\]
by taking thermal averages, we find
\[
\la a_i^\dg a_j^\dg a_k a_l\ra=\frac{e^{\b\o_i}-q}
{q^{1-\d_{ik}-\d_{il}}e^{\b\o_i}-q}(
\la a_j^\dg a_l \ra\la a_i^\dg a_k\ra+q^{-\d_{ij}}
\la a_j^\dg a_k \ra\la a_i^\dg a_l \ra) \ .
\]
With coinciding modes\footnote
{
Recall that ${\bf K}\equiv\frac{1}{2}({\bf k}_a+{\bf k}_b),\ $
${\bf p}\equiv{\bf k}_a-{\bf k}_b$; in the case ${\bf p}=0,\ $
${\bf K}={\bf k}_a={\bf k}_b$. }
, this leads to the formula
\beq
\la a_i^\dg a_i^\dg a_i a_i\ra=\frac{1+q}
{(e^{\b\o_i}-q)(e^{\b\o_i}-q^2)}\ .\ \
\eeq
{}From the last relation and distribution (22),
the ratio under question follows:
\beq
\tilde{\l}_i=\frac{\la a_i^\dg a_i^\dg a_i a_i\ra}
{\la a_i^\dg a_i \ra^2}=
\frac{(1+q)(e^{\b\o_i}-q)}
{e^{\b\o_i}-q^2} .
\eeq
This constitutes one of our main results.
For convenience, let us set
\beq
\tilde{\l}=1+\l \ \ \ \ {\rm with} \ \ \ \
\l=q\frac{e^{\b\o}-1}{e^{\b\o}-q^2}\ .
\eeq
The quantity $\l$ can be directly confronted with
empirical data. Note that in the non-deformed limit $q\to 1$
the value $\l_{\rm BE}=1$,
proper for Bose-Einstein statistics, is
correctly reproduced from Eq.~(28). This, obviously,
corresponds to the Bose-Einstein distribution contained
in (22) if $q\to 1$.

\vspace{-0.1cm}

\begin{figure} [ht]
\begin{center}
\epsfig{file=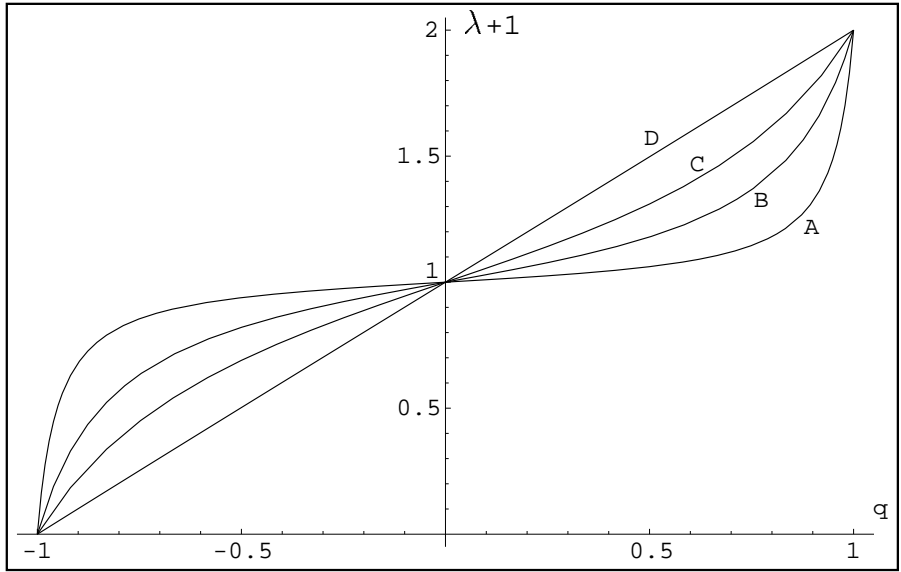,width= 12cm,angle=0}
\vspace{0.6cm}
\caption{   Intercept $\l $ versus deformation parameter $q$,
as given by Eq.~(28). The curves A, B, C and D correspond to
the values $w_A= 0.1,\ $ $w_B=0.35,\ $ $w_C=0.8\ $ and $w_D=5.0\ $
of the dimensionless variable $w\equiv \b \o$.}
\end{center}
\end{figure}

\vspace{-2mm}

At $q=-1$, Eqs.~(22) and (28) {\it formally
coincide} with the Fermi-Dirac distribution
and the value $\l_{\rm FD}=-1$ proper for
the Fermi-Dirac statistics respectively, although the defining
relations are not exactly those of the fermionic system
(rather, the hard core
bosons \cite{Sa}).
Finally, at $q=0$ we get $\l = 0$,\ which coincides with the
analogous fact for the case of purely classical description
(complete absence of quantum effects due to identical particles).
The three different cases are clearly seen in Fig.~2 as
the only three points where all the different curves (the continuum
parametrized by $w=\b \o$) merge and, thus,
the dependence on momentum and/or temperature disappears. From
the continuum of curves, there exists a {\bf unique limiting}
(asymptotic) one $\tilde{\l}=1+q\ $ (or $\l=q$), which corresponds
to the limit $w\to\infty$ (i.e. to zero
temperature or infinite momentum).   Conversely, for very large
temperature such that $w\to 0$, the curve
goes over into the step-shaped function
\[
\tilde\l=\cases{ 0, \ \ \ \ \ \ q=-1,\cr
          1, \ \ \  -1< q < 1, \cr
          2, \ \ \ \ \ \ q=1 ,}
\]
with the constant $\l=0$ for each fixed value of $q$ except for
the endpoints $q=1$ and $q=-1$.

We find now the formula describing two-particle correlations
(at identical momenta), which corresponds to Biedenharn-Macfarlane
$q$-oscillators, see Eq. (16). From the relation
\[
\la b_i^\dg b_i^\dg b_i b_i\ra -q^{2} \la b_i^\dg b_i b_i b_i^\dg\ra
=-\la b_i^\dg b_i q^{N_i} \ra (1+q^2) \, ,
\]
which is valid just for coinciding modes (i.e. equal momenta), we
immediately get
\[
\la b_i^\dg b_i^\dg b_i b_i \ra=\frac{1+q^2}{q^2 e^{\b\o_i}-1}
\la b_i^\dg b_i q^{N_i}\ra  \ .
\]
The thermal average
in the r.h.s. can be easily evaluated to yield
$\la b_i^\dg b_i q^{N_i}\ra = q/(e^{\b\o_i}-q^2) .$
Taking this into account, we find the expression for
two-particle distribution, namely
\beq
\la b_i^\dg b_i^\dg b_i b_i\ra=\frac{2\cos\theta}
{e^{2\b\o_i}-2\cos(2\theta)e^{\b\o_i}+1}\ .
\eeq
{}From that, the desired formula for (the intercept of)
two-particle correlations finally results in:
\beq
\tilde\l_i\equiv\l_i + 1 =\frac{\la b_i^\dg b_i^\dg b_i b_i\ra}
{(\la b_i^\dg b_i \ra)^2}=
\frac{2\cos\theta(t_i+1-\cos\theta)^2}{t_i^2+2(1-\cos^2\theta)t_i}\ ,
\eeq
where $t_i=\cosh(\b\o_i)-1$.
Note that both (29) and (30) are real functions (as they should)
since, like (25), they both depend on the complex $q$-parameter
(19) through the combination $\frac12 (q+q^{-1})=\cos\t $.

In the rest of this section, we extract some useful
information contained in Eq.~(30).
Solving this equation with respect to $\cos\t$ at a fixed value
$\l=\bar{\l}$, we obtain the deformatin angle as the function:
$\t=\t(\bar{\l},\ {\bf K},\ T,\ m)$.

Figure 3 is used to illustrate main properties of the intercept
(correlation strength) $\l$ treated from the standpoint of $q$-deformation,
that is, on the base of Eq.~(30).
First, let us note that the continuum of curves
$\tilde\l=\tilde\l (\cos\t )$ parametrized by $w=\b\o $ divides
into three different classes (``subcontinua'') given by three
intervals of the parameter:\ (i) $0<w\le w_0 $, (ii) $w_0<w<w'_0 $,
and (iii) $w'_0\le w<\infty $.
Here, the two ``critical'' values
$w_0=w_B\simeq 0.481\ $ and $w'_0=w_D\simeq 0.696\ $
are singled out (curves B and D respectively).
The curves A, C, and E are typical representatives of the
classes (i),(ii) and (iii). All the curves from classes (i), (ii)
possess two extrema, the minimum being to the left of the maximum
(the curve D is unique since its extrema degenerate, coinciding
with the point of inflection).
This fact enables us to
define naturally ``the range of small deformations'' -- the interval
$I_{\rm small}$ for the variable $\t$: from $\t =0$
(no deformation) to the value
yielding minimal $\l $, $\l_{\rm min}\approx 0.33$, implied by
the ``critical'' value $w_0=w_D$.
It is seen that on the interval
$I_{\rm small}$ the intercept $\l $ monotonically decreases with
increasing of $\t$ (or $1-\cos\t $), the strength of deformation.
On the contrary, each curve from the third class
is monotonic as a whole and, thus, there is no criterion
(no peculiar point), which would naturally separate
``small'' deformations from ``large'' ones.

\begin{figure}[t]
\begin{center}
\vspace{-0.6cm}
\epsfig{file=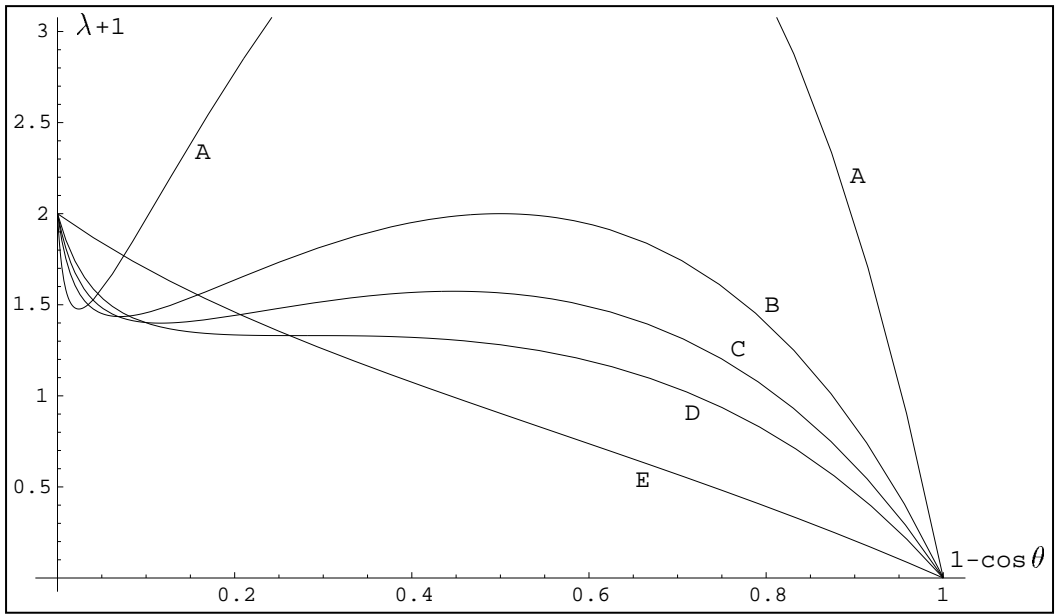,width=14cm,angle=0}
\vspace{0.1cm}
\caption  {  Intercept $\l $ versus deformation given by $\cos\t$,
see Eq.~(30). The curves A,B,C,D and E correspond to
the values $w_A= 0.3,\ $ $w_B=0.481,\ $ $w_C=0.58,\ $
$w_D=0.696\ $and $w_E=2.0\ $
of the dimensionless variable $w\equiv \b \o$.  }
\vspace{-0.4cm}
\end{center}
\end{figure}

In this paper we deal only with classes (ii) and (iii) since
all their curves at $\t\ne 0$ lie below the straight line
$\tilde\l = 2\ $ -- the largest possible correlation
attainable in the Bose-Einstein
case (note that curve B contains, besides $\t =0 $, just a
{\it single} point at a certain value of $1-\cos\t$, where
the value $\tilde\l = 2\ $ is also attained).
Moreover,
for the (ii)-type curves, we restrict ourselves to
$I_{\rm small}$, ignoring ``moderately large'' deformations
(between min. and max.), for which the behaviour of
$\l=\l (\t)$ is opposite to that for $I_{\rm small}$,
as well as very large ones (to the right of the maximum).
Unlike these two regular classes, the class (i) consists of
``irregular'' curves: for each such curve there exist
$q$-deformations that generate correlation strengths
{\it exceeding the maximal possible one} $\tilde\l = 2\ $.
Therefore, we discard the class (i), at least at this stage.

Finally, let us discuss special values of the physical variables
$T,\ |{\bf K}|$, which provide the peculiar values
$w_0=w_B\simeq 0.481 $ and  $w'_0=w_D\simeq 0.696 $ (recall that
$w=\sqrt{m^2 + {\bf K}^2}/T$).
With $m(\pi^{\pm,0})=139.57$ MeV
and lowest mean momentum of the pion pair fixed to be $|{\bf K}| = 0$,
we get
two bounded from below values for the temperature: $T_0=290.0$~MeV
and $T'_0=200.5$~MeV.
It is interesting to compare these data with
that for the typical curve from class (iii). Namely, at $w=w_E=2.0 $
(the curve E) for pions of this same lowest momentum
we get the limiting temperature $T_E=69.8$ MeV.


\section{ Discussion and outlook }
\label{sec5}


The main purpose of theoretical approaches to RHIC is to find an adequate
description for the non-equilibrium state formed during the collision.
On this way, the $q$-boson techniques enables us to treat the
non-stationary hot and dense matter effectively
as a ``noninteracting ideal gas''.
To deal with $q$-bosons, it is necessary to determine the $q$-parameter that
corresponds to the actual state of the hot medium.
We propose a way of extracting from
the two-particle corre\-la\-tions a useful
information concerning $q$, and develop an effective picture of the
two-pion (-kaon) spectra in RHIC. According to our results,
the measured deviation from unity
of the intercept $\l $ is interpreted
as the manifestation of $q$-boson properties of
the pion system created in RHIC.

From Eqs.~(28) and (30), one can express the quantities encoding the deformation
as: $q=q(\l,\ {\bf K},\ T,\ m)$ and $\t=\t(\l,\ {\bf K},\ T,\ m)$.
In the characteristic limits of temperature and momentum, our model exhibits
a remarkable feature, valid for both types of exploited $q$-oscillators:

$\bullet$ {\it For very low temperature at fixed momenta,
or very large momenta at fixed temperature}
(i.e. at $w\to\infty ,$ compare with curve D in Fig.~2 and curve E in Fig.~3),
{\it we come to the equality}
\beq
  \hspace{26mm}  q=\l   \hspace{28mm}
      (T\to 0 \ \ \ \ {\rm or} \ \ \  \vert{\bf K}\vert \to\infty)
\eeq
{\it with the type ``A'' oscillators, and to the equality}
\beq
   \hspace{26mm} 2\cos\t=\l+1 \hspace{14mm}
      (T\to 0 \ \ \ \ {\rm or} \ \ \  \vert{\bf K}\vert \to\infty)
\eeq
{\it for the type ``B'' oscillators.} This implies the unified
{\bf direct} connection $\l\leftrightarrow q$, namely
$\tilde\l = [2], $ for both types ``A'' and  ``B''.

On the other hand, finite temperature and momenta become non-trivially
involved (especially in the case of type ``B'' $q$-bosons)
in the relation between $\l$ and the deformation
parameter $q$, see Eqs. (28), (30) as well as Figs.~2,~3.

$\bullet$
Equation (30) implies {\it dependence of the intercept $\l$ on the pair
mean momentum} ${\bf K}$ for fixed values of the deformation angle
$\t$, temperature and particle mass, since $\l=\l(\t,\ {\bf K},\ T,\ m)$.
In Fig.~4, we present this dependence exemplifying it, for $T=120$ MeV
and $m=m_\pi$, with four curves which correspond to the values
$4^{\circ}$, $9^{\circ}$, $15^{\circ}$ and $24^{\circ}$
of the deformation angle $\theta$. Each curve tends to its own
asymptote given by Eq.~(32). The apparent {\it variability}
of $\lambda$ with varying $\vert{\bf K}\vert$ (for a fixed deformation)
is one of the consequences of our model, and can be viewed both as
an interesting prediction and as a check-point for this approach.

\begin{figure}[ht]
\begin{center}
\vspace{-0.6cm}
\epsfig{file=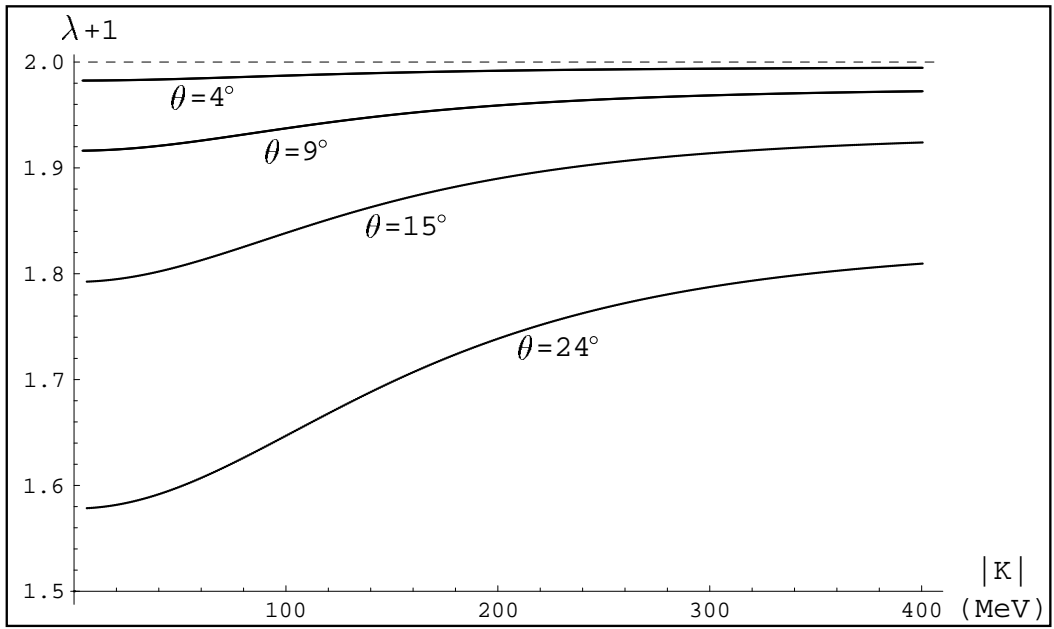, width=14cm,angle=0}
\vspace{0.1cm}
\caption  {    Dependence of the intercept $\lambda$ on the pion
momentum $|{\bf K}|$ for some values of $\theta $. The temperature
is fixed to be  $T=120$ MeV; $\ m=m_\pi$.   }
\vspace{-0.4cm}
\end{center}
\end{figure}

In \cite{GM}, it was argued that the {\it quon-based} description
applied to pions participating in the decay $K_{\rm L}\to \pi^+ \pi^-$,
can give a complete account of this process
(instead of $CP$-violation, as commonly accepted cause).
Moreover, this sets a bound on
the ``strength'' of the deformation, namely $\ 1-q \le 10^{-6}$.
On the contrary, within the proposed approach to
multiparticle correlations (as occurred
in the HBT interferometry) based on viewing
the identical pions as $q$-bosons,
the two-pion correlations imply the employment of
significantly more developed
(strength of) $q$-deformation. Namely, in our
implementation of the $q$-Bose gas picture,
the empirical data are interpreted as corresponding to the strength
$\, \simeq 0.2 --\, 0.6\, $ of $q$-deformation.

The approach proposed in this paper, we hope, opens interesting new
perspectives for further research in this direction.
Three-particle correlations as well as
fermion-fermion(-fermion) ones are the topics
to be studied next using the developed ``$q$-techniques''.

\vspace{.8cm}

\noindent
{\large{\bf Acknowledgements }}
\label{sec6}


\vspace{.2cm}
\noindent
D.A. acknowledges many helpful and instructive discussions with U.~Heinz.
The work of two of us (A.G. and N.I.) was partially supported by
the Award No. UP1-309 of the U.S. Research and Development Foundation
for the Independent States of the Former Soviet Union (CRDF), and by
the Ukrainian DFFD Grant 1.4/206.


\section*{APPENDIX}


In the appendix we consider the two-particle quantum statistical
correlations  when the final state interactions of the
detected particles are neglected.
This phenomenon is visualized most transparently
on the basis of standard quantum mechanics in the non-relativistic case.
Meanwhile, using the presented scheme, the relativistic picture
can also be considered.
Moreover, this approach allows us to take
into consideration the final state interactions as well \cite{anch98}.

The probability to register two particles,
with definite asymptotic momenta ${\bf p}_a$ and ${\bf p}_b$, which are
created in the relativistic heavy-ion collisions is usually compared
with the probability to register two particles of
the same momentum independently.
For that reason, we begin by considering the single-particle spectrum.

\section*{1. Single-particle cross section}

Let us consider the single-particle state $\psi_\gamma$ of a particle
emitted by the source. Its propagation to the detector is governed by the
Schr\"odinger equation
 \begin{equation}
  i \frac{\partial \psi_\gamma({\bf x},t)}
         {\partial t} =
  \hat{h}({\bf x})\,
  \psi_\gamma({\bf x},t) \, ,
 \label{a101}
\hspace{8mm}  {\rm where}  \hspace{8mm}
  \hat{h} ({\bf x}) = - \frac{1}{2m} {\bf \nabla}^2 \, .
 \label{a102}
 \end{equation}
The index $\gamma$ denotes a complete set of 1-particle quantum numbers.
(In a basis of wave packets, these could contain the
centers ${\bf X}$ of the wave packets of the
particles at their freeze-out times $t$.)
Equation (\ref{a101}) is solved by
 \begin{equation}
  \psi_\gamma({\bf x},t,t_0) =
  e^{  - i \hat{h}({\bf x}) (t-t_0) }\,
  \psi_\gamma({\bf x},t_0)
\, ,
 \label{a103}
 \end{equation}
in terms of the single-particle wave function at some initial time $t_0$.
We will assume that the detector measures asymptotic momentum
eigenstates, i.e. that it acts by projecting the emitted 1-particle
state onto
 \begin{equation}
  \phi^{\rm out}_{\bf p}({\bf x},t)
  = e^{i ({\bf p}\cdot {\bf x} - \omega ({\bf p}) t)} \ ,
 \label{a104}
 \end{equation}
where $\omega ({\bf p}) =  {\bf p}^2/2m$
is the energy of the particle.
The measured single-particle momentum amplitude is then
 \begin{equation}
   A_\gamma({\bf p},t_0) = \lim_{t\to\infty}
   \int d^3x\,
   \phi^{\rm out,*}_{\bf p} ({\bf x},t) \,
   \psi_\gamma({\bf x},t,t_0) \, .
 \label{a105}
 \end{equation}
Using the time evolution equation (\ref{a103}), this can be expressed in
terms of the emitted single-particle wave function $\psi_\gamma$
at earlier times as
\begin{eqnarray}
   A_\gamma({\bf p},t_0) &=& \lim_{t\rightarrow \infty}
   \int d^3x \,
   \left[ e^{ - i \hat{h}({\bf x}) (t_0-t)} \,
          \phi^{\rm out}_{\bf p} ({\bf x},t)
   \right]^* \,
   \psi_\gamma({\bf x},t_0)
\nonumber \\
&=&
   \int d^3x \,
  e^{i \omega ({\bf p}) \, t_0 - i{\bf p}\cdot {\bf x}} \,
   \psi_\gamma({\bf x},t_0)
   \, .
 \label{a106}
\end{eqnarray}
This expression means that the momentum amplitude, as it should be, is
an on-shell Fourier transformation of the emitted wave at emission times.
It is worth noting that this is not the case when, after emission, a
particle is subject to final state interactions.

The single-particle probability is obtained by averaging (\ref{a106})
and its complex conjugate using the density matrix defining the source.
This density matrix is characterized by a probability distribution for
the single-particle quantum numbers $\gamma $.
We write
 \begin{equation}
   P_1({\bf p})  =
   \overline{
   \Big\langle \big|
   A_\gamma({\bf p},t_0)
   \big|^2 \Big\rangle} _\gamma
     =
   \ \,
   \sum _{\gamma \gamma '} \int dt_0\, dt_0'\,
    \rho _{\gamma ,\gamma '}\,
   A_{\gamma '} \left( {\bf p},t_0' \right)\,
   A^*_{\gamma } \left( {\bf p},t_0 \right)
   \, ,
 \label{a107}
 \end{equation}
where the
anglular brackets mean averaging over source quantum numbers $\gamma$;
the overline means averaging over emission times and is detailed in
the r.h.s.
Inserting (\ref{a106}) into (\ref{a107}) yields
\begin{equation}
    P_1({\bf p}) =
   \int d^4x\, d^4x'\,
   e^{ip\cdot (x-x')} \,
   \sum_{\gamma , \gamma '} \rho_{\gamma \gamma '} \,
   \psi _\gamma (x) \,
   \psi^*_{\gamma '}(x')
   \, ,
 \label{a108}
\end{equation}
where $p=(\omega ({\bf p}),{\bf p})$.
Let us introduce new time and position space variables
\begin{equation}
    Y=\frac{1}{2} (x+x') \ , \ \ \ \ \ \ y=x-x' \ .
 \label{a109}
\end{equation}
We now define the single particle Wigner density $S(X,K)$ of the
source as
 \begin{equation}
  S(Y,K) = \int d^4y\, e^{i K\cdot y}\,
  \sum_{\gamma , \gamma '} \rho_{\gamma \gamma '}\,
  \psi_\gamma \left(Y+{\textstyle{y\over 2}}\right) \,
  \psi_{\gamma '}^*\left(Y-{\textstyle{y\over 2}}\right) \, .
 \label{a110}
 \end{equation}
Using the hermiticity of the source density matrix $\rho_{\gamma \gamma '}$
it is  easily shown that $S(Y,K)$ is real.
Thus, we come to the expression for the single-particle
spectrum, which employs the source function:
 \begin{equation}
   P_1({\bf p}) =   \int d^4x\, S(x,p)
  \,
 \label{a112}
 \end{equation}
where the integration goes over emission times.

\section*{2. Two-particle quantum statistical correlations without final
state interactions}

Let us consider a two-particle state $\psi_\gamma$ emitted by the
source. Its propagation to the detector is governed by the
Schr\"odinger equation
 \begin{equation}
  i \frac{\partial \psi_\gamma({\bf x}_a,{\bf x}_b,t)}
         {\partial t} =
  \hat{H}({\bf x}_a,{\bf x}_b)\,
  \psi_\gamma({\bf x}_a,{\bf x}_b,t) \, ,
 \label{a3}
 \end{equation}
where
$\hat{H}({\bf x}_a,{\bf x}_b) = \hat{h} ({\bf x}_a) + \hat{h} ({\bf x}_b)$.
The index $\gamma$ denotes a complete set of 2-particle quantum numbers.
(In a basis of products of two wave packets these could contain the
centers ${\bf X}_a$, ${\bf X}_b$ of the wave packets of the two
particles at their freeze-out times $t_a$, $t_b$, respectively.)
Equation (\ref{a3}) is solved by
 \begin{equation}
  \psi_\gamma({\bf x}_a,{\bf x}_b,t) =
  e^{- i \hat{H}({\bf x}_a,{\bf x}_b) (t-t_0)}\,
  \psi_\gamma({\bf x}_a,{\bf x}_b,t_0)
 \label{a6}
 \end{equation}
in terms of the two-particle wave function at some initial time $t_0$.
We will assume that the detector measures asymptotic momentum
eigenstates, i.e. that it acts by projecting the emitted 2-particle
state onto
 \begin{equation}
  \phi^{\rm out}_{{\bf p}_a,{\bf p}_b}({\bf x}_a,{\bf x}_b,t)
  = e^{i ({\bf p}_a\cdot {\bf x}_a - \omega_a t)}\,
    e^{i ({\bf p}_b\cdot {\bf x}_b - \omega_b t)}
\, ,
 \label{a7}
 \end{equation}
where
 $\omega_{a,b} = {\bf p}_{a,b}^2/2m^2$.
We will only consider
the case of pairs of identical particles, $m_a=m_b=m$.
The measured two-particle momentum amplitude is then
 \begin{equation}
   A_\gamma({\bf p}_a,{\bf p}_b) =
   \lim_{t\to\infty}
   \int d^3x_a\, d^3x_b\,
   \phi^{\rm out,*}_{{\bf p}_a,{\bf p}_b} ({\bf x}_a,{\bf x}_b,t) \,
   \psi_\gamma({\bf x}_a,{\bf x}_b,t) \, .
 \label{a8}
 \end{equation}
If we consider identical particles, then the two-particle wave function
$\psi_\gamma({\bf x}_{a},{\bf x}_{b},t)$
should be symmetrized (antisymmetrized). Writing this explicitly, we obtain
 \begin{equation}
   A_\gamma({\bf p}_a,{\bf p}_b) =
   \lim_{t\to\infty}
   \frac{1}{\sqrt{2} } \int d^3x_a\, d^3x_b\,
   \phi^{\rm out,*}_{{\bf p}_a,{\bf p}_b} ({\bf x}_a,{\bf x}_b,t) \,
    \left[ \psi_\gamma({\bf x}_a,{\bf x}_b,t) +
          e^{i\alpha }\, \psi_\gamma({\bf x}_b,{\bf x}_a,t)\right]
\, ,
 \label{a8a}
 \end{equation}
where $\alpha =0$ ($\alpha =\pi $) for identical bosons (fermions).
Relabelling the variables of integration in the second term on the r.h.s.
of this equation as
${\bf x}_a \to {\bf x}_b$ and ${\bf x}_b \to {\bf x}_a$,
we come to the expression:
 \begin{eqnarray}
   A_\gamma({\bf p}_a,{\bf p}_b) =
   \lim_{t\to\infty}
   \frac{1}{\sqrt{2} } \int d^3x_a\, d^3x_b\, &&
  \! \! \! \! \! \! \! \! \!
     \left[ e^{i ({\bf p}_a\cdot {\bf x}_a - \omega_a t)}\,
           e^{i ({\bf p}_b\cdot {\bf x}_b - \omega_b t)}\, + \,
      e^{i\alpha }\, e^{i ({\bf p}_a\cdot {\bf x}_b - \omega_a t)}\,
                     e^{i ({\bf p}_b\cdot {\bf x}_a - \omega_b t)}
      \right]^*
 \nonumber\\
&& \! \! \! \! \! \!   \times \   \psi_\gamma({\bf x}_a,{\bf x}_b,t)
\, .
 \label{a8b}
 \end{eqnarray}
This simple algebra results in the evident conclusion that it does not matter
which wave function should be symmetrized (antisymmetrized) at asymptotic
time $t$: two-particle wave function
$\psi_\gamma({\bf x}_{a},{\bf x}_{b},t)$ or out-state wave function.
We show below that the same holds for emission times.

Using the evolution operator (see (\ref{a6})), Eq.~(\ref{a8}) can be
expressed in
terms of the two-particle wave function $\psi_\gamma$ emitted at
earlier times as
 \begin{equation}
  A_\gamma({\bf p}_a,{\bf p}_b) = \lim_{t\rightarrow \infty}
   \int d^3x_a\, d^3x_b\,
   \left[\, e^{ - i \hat{H}({\bf x}_a,{\bf x}_b) (t_0-t)} \,
          \phi^{\rm out}_{{\bf p}_a,{\bf p}_b} ({\bf x}_a,{\bf x}_b,t)
   \right]^* \,
   \psi_\gamma({\bf x}_a,{\bf x}_b,t_0)
\, .
 \label{a9}
 \end{equation}
This is correct for all times $t_0 \ge {\rm max}\, [t_a,t_b]$, where
$t_{a,b}$ are the freeze-out times for the two particles. Note that
the time evolution operator has been shifted
(by Hermitian inversion of the unitary evolution operator)
from the emitted
two-particle state with arbitrary quantum numbers $\gamma$
to the two-particle momentum eigenstate
$\phi^{\rm out}_{{\bf p}_a,{\bf p}_b} ({\bf x}_a,{\bf x}_b,t)$
which is thereby transformed into a plane wave at time $t_0$.
It should be pointed, that such an evolution
of the out-state from $t=\infty $ to $t=t_0$ if one includes the
two-particle interaction into the
Hamiltonian $\hat{H}({\bf x}_a,{\bf x}_b)$,
would bring a distorted wave at time $t_0$ instead of the plane one
\cite{anch98}.

We assume that the two particles are emitted independently, implying that,
at some freeze-out time $t_a$, the two-particle wave function
$\psi_\gamma({\bf x}_{a},{\bf x}_{b},t)$
factorizes
 \begin{equation}
   \psi_\gamma({\bf x}_a,{\bf x}_b,t_a) = \frac{1}{\sqrt{2}}
   \left[ \psi_{\gamma_a}({\bf x}_a,t_a)\,
          \psi_{\gamma_b}({\bf x}_b,t_a)
   + \, e^{i \alpha}
          \psi_{\gamma_a}({\bf x}_b,t_a)\,
          \psi_{\gamma_b} ({\bf x}_a,t_a) \right]
\, ,
 \label{a10}
 \end{equation}
The indices $\gamma_a,\gamma_b$ on the 1-particle wave functions now
label complete sets of 1-particle quantum numbers.
Time $t_a$ is the emission time of the latest emitted particle.
At this time the first emitted particle of the pair has already
propagated for a time $t_a-t_b$ if it was emitted at $t_b <
t_a$. During this time the first emitted particle cannot ``see'' the
second particle as a separate entity, but only as part of the remaining
fireball. That is why the two-particle symmetrization (antisymmetrization)
can be effectively extracted from many-particle symmetrization
(antisymmetrization) after emission of the second particle.
In other
words a factorization of the two-particle wave function from the
many-particle wave function, which corresponds to the total system, is
possible only when the second emitted particle is frozen out from the
fireball.
This means that we adopt concept of the two-particle
amplitude factorization.
All these considerations complicate when we start with the
two-particle Hamiltonian that contains terms where the coordinates are
entagled, for
instance the two-particle potential energy $V({\bf x}_a-{\bf x}_b)$; but
a discussion of this point is beyond the scope of the present letter
(see \cite{anch98} for details).

Owing to the particle-particle emission symmetry and to the commutation
of the free evolution operators, we can write the
measured two-particle momentum amplitude as
 \begin{eqnarray}
   A_{\gamma_a, \gamma_b}({\bf p}_a,{\bf p}_b,t_a,t_b) =
   \lim_{t\to\infty} && \! \! \! \! \! \!
   \frac{1}{\sqrt{2}} \int d^3x_a\, d^3x_b\,
   \phi^{\rm out,*}_{{\bf p}_a,{\bf p}_b} ({\bf x}_a,{\bf x}_b,t) \,
 \nonumber \\
&& \! \! \! \! \! \! \times \, \big[
  e^{- i[\hat{h} ({\bf x}_a) (t-t_a)+\hat{h} ({\bf x}_b) (t-t_b)]}\,
      \psi_{\gamma_a}({\bf x}_a,t_a)\, \psi_{\gamma_b}({\bf x}_b,t_b) \, +
 \nonumber \\
&&    + \ e^{i \alpha}
  e^{- i[\hat{h} ({\bf x}_b) (t-t_a)+\hat{h} ({\bf x}_a) (t-t_b)]}\,
      \psi_{\gamma_a}({\bf x}_b,t_a)\, \psi_{\gamma_b}({\bf x}_a,t_b) \big]
\, .
 \label{a17}
 \end{eqnarray}
Now, we first
relabel the variables of integration in the second term on the r.h.s.
of this equation, what results in symmetrization (antisymmetrization) of the
out-state.
Secondly, we invert the evolution operator, which therefore acts on
the out-state
and brings it from asymptotic time $t$ to initial times $t_a$ and $t_b$.
As a result, we come to the final expression of the two-particle amplitude:
\begin{equation}
   A_{\gamma_a, \gamma_b}({\bf p}_a,{\bf p}_b,x_a^0,x_b^0) =
  \frac{1}{\sqrt{2} } \int d^3x_a\, d^3x_b \,
\left[
      e^{i(p_a\cdot x_a + p_b\cdot x_b) }
      + e^{-i\alpha }\,
      e^{ i(p_a\cdot x_b + p_b\cdot x_a) }
\right]
\,
      \psi_{\gamma_a}(x_a)\, \psi_{\gamma_b}(x_b)
\ ,
\label{a19}
\end{equation}
where $x_a^0=t_a\ \  {\rm and}\ \ x_b^0=t_b$.
We therefore represent the measured two-particle momentum amplitude
as a projection of a non-symmetrized two-particle wave function taken at
emission times onto symmetrized (antisymmetrized) plane waves taken
at emission times as well.

\medskip

The two-particle probability is obtained by averaging (\ref{a19})
and its complex conjugate with the density matrix defining the source.
This density matrix is characterized by a probability distribution for
the two-particle quantum numbers $(\gamma_a,\gamma_b)$ and by a
distribution of emission times $(t_a,\, t_b)$. We write:
 \begin{eqnarray}
   \! \! \! \! \! P_2({\bf p}_a,{\bf p}_b) \! \! &=& \! \!
   \overline{
   \Big\langle \big|
   A_{\gamma_a,\gamma_b}({\bf p}_a,{\bf p}_b;t_a,t_b)
   \big|^2 \Big\rangle} _{\gamma_a,\gamma_b} =
   \\
   &=& \! \!
    \sum  _{\gamma_a \gamma_b,\gamma_{a'} \gamma_{b'} }
   \int dt_a\, dt_b\, dt_{a'}\, dt_{b'}\,
   \rho _{\gamma_a \gamma_{a'}}\, \rho _{\gamma_b \gamma_{b'}}\,
   A_{\gamma_{a'} \gamma_{b'} }
                   \left( {\bf p}_a,{\bf p}_b;t_{a'},t_{b'} \right)\,
   A^*_{\gamma_a \gamma_b } \left({\bf p}_a,{\bf p}_b ;t_a,t_b\right)
\nonumber \, ,
 \label{a20}
 \end{eqnarray}
where anglular brackets on the r.h.s. of this equation
mean averaging over the quantum numbers $\gamma_a, \, \gamma_b$
and the overline means averaging over the initial (emitting) times.
We made the ansatz
$\rho _{\gamma_a \gamma_b,\gamma_{a'} \gamma_{b'} } =
   \rho _{\gamma_a \gamma_{a'}}\,
   \rho _{\gamma_b \gamma_{b'}}$,
which factorizes the initial density matrix
$\rho _{\gamma_a \gamma_b,\gamma_{a'} \gamma_{b'} }$
in such a way that independent emission of the
two particles is ensured.

According to (\ref{a19}), the probability consists of
four terms, which we write as
 \begin{equation}
   P_2({\bf p}_a,{\bf p}_b) =
   P_{11} + P_{22} + P_{12} + P_{21} \, ,
\label{a22}
\end{equation}
which have the structure
$(a_1+a_2)(a_1+a_2)^*=a_1a^*_1+a_2a^*_2+a_1a^*_2+a_2a^*_1$.
First diagonal term reads
\begin{eqnarray}
P_{11}(p_a,p_b) =
&& \frac{1}{2} \int d^4x_a\, d^4y_a \,
e^{i  (p_a\cdot x_a - p_a\cdot y_a) } \,
\Big\langle
 \psi _{\gamma_a}(x_a)\, \psi ^*_{\gamma_a}(y_a)
 \Big\rangle _{\gamma_a}
\nonumber \\
&& \times
\int d^4x_b \, d^4y_b\,
e^{i  (p_b\cdot x_b - p_b\cdot y_b) } \,
\Big\langle
 \psi _{\gamma_b}(x_b)\, \psi ^*_{\gamma_b}(y_b)
 \Big\rangle _{\gamma_b}
\ .
\label{a23}
\end{eqnarray}
We introduce the new momentum variables $K$ and $p$ as
\begin{equation}
K=\frac{1}{2} (p_a+p_b) \ , \ \ p=p_a-p_b \ \ \
\Rightarrow \ \ \
p_a=K+\frac{p}{2}  \ , \ \ p_b=K-\frac{p}{2} \ .
\label{a22a}
\end{equation}
Using these variables and the space-time variables (\ref{a109}),
$P_{11}$ can be rewritten in the form
\begin{eqnarray}
 \! \! P_{11}(p,K) = && \! \! \!
\frac{1}{2}
\int d^4X\, \int d^4x \,
e^{\imath  (K+\frac{1}{2} p)\cdot x } \,
\Big\langle
\psi _{\gamma_a}\Big( X+\frac{x}{2} \Big)\,
\psi ^*_{\gamma_a}\Big( X-\frac{x}{2} \Big)
 \Big\rangle _{\gamma_a}
\nonumber \\
&& \! \! \!
\times \,
\int d^4Y\, \int d^4y \,
e^{\imath  (K-\frac{1}{2} p)\cdot y } \,
\Big\langle
\psi _{\gamma_b}\Big( Y+\frac{y}{2} \Big)
\psi ^*_{\gamma_b}\Big( Y-\frac{y}{2} \Big)
 \Big\rangle _{\gamma_b}
\ .
\label{a24}
\end{eqnarray}
In the integrals over the variables $x$ and $y$, we immediately recognize
the
freeze-out Wigner density (\ref{a110}). But before rewriting this term
let us note that $P_{22}$
can be obtained from $P_{11}$ by a mutual
change of momenta $p_a\rightleftharpoons  p_b$, which in turn results in
changing $p$ to $-p$.
Hence, to obtain
$P_{22}$ from $P_{11}$ we need only to change the sign before $p$.
Performing such a change, we come to an equality of the diagonal terms
$P_{11}(p,K) = P_{22}(p,K)$. Putting together these two terms, we obtain
\begin{equation}
P_{11}(p,K)+ P_{22}(p,K) =
\int  d^4 X \, S\Big( X,K+\frac{p}{2} \Big) \,
\int  d^4 Y \, S\Big( Y,K-\frac{p}{2} \Big)
\ ,
\label{a28}
\end{equation}
i.e. the product of two single-particle probabilities
(see Eq.~(\ref{a112}))
to registrate independently two particles with asymptotic momenta
${\bf p}_a={\bf K}+{\bf p}/2$ and
${\bf p}_b={\bf K}-{\bf p}/2$, respectively.

We now turn to considering of the cross term $P_{12}=a_1a_2^*$, which
is a complex conjugate to the second cross-contribution $P_{21}=a_1^*a_2$,
hence their sum is real. We have
\begin{eqnarray}
P_{12}(p_a,p_b) =
&& \frac{1}{2} \, e^{i\alpha}\, \int d^4x_a\, d^4y_a \,
e^{i  (p_a\cdot x_a - p_b\cdot y_a) } \,
\Big\langle
 \psi _{\gamma_a}(x_a)\, \psi ^*_{\gamma_a}(y_a)
 \Big\rangle _{\gamma_a}
\nonumber \\
&& \times
\int d^4x_b \, d^4y_b\,
e^{i  (p_b\cdot x_b - p_a\cdot y_b) } \,
\Big\langle
 \psi _{\gamma_b}(x_b)\, \psi ^*_{\gamma_b}(y_b)
 \Big\rangle _{\gamma_b}
\ .
\label{a29}
\end{eqnarray}
By the same change of variables as for diagonal terms, the exponents
in the integrand can be rewritten as:
$(p_a\cdot x_a - p_b\cdot y_a) = (p\cdot X + K\cdot x)$
and
$(p_b\cdot x_b - p_a\cdot y_b) = (-p\cdot Y+ K\cdot y)$.
Using the definition of the source function (\ref{a110}), we obtain:
\begin{equation}
P_{12}(p,K) =
\frac{1}{2} \, e^{i\alpha}\, \int  d^4X \, e^{i  p\cdot X }\, S(X,K)\,
\int  d^4Y \, e^{-i  p\cdot Y }\, S(Y,K) \ .
\label{a30}
\end{equation}
The second cross term $P_{21}$, which is complex conjugate to $P_{12}$, is
proportional to the exponent $\exp{(-i\alpha)}$, but the remaining factor
is real.
Hence,  the sum of two cross terms looks like
$P_{12}+P_{21}=\frac{1}{2} Ae^{i\alpha}+\frac{1}{2} Ae^{-i\alpha}=
A\cos{\alpha} $, where $A$ is real.
We are now ready to write the total expression for the two-particle
probability
\begin{eqnarray}
P_2(p,K) =&&
\int  d^4 X  S\Big( X,K+\frac{p}{2} \Big)
\int  d^4 Y  S\Big( Y,K-\frac{p}{2} \Big)
\nonumber \\
+&& \cos{\alpha} \,
\int  d^4 X \,  e^{i  p\cdot X } S(X,K)
\int  d^4 Y  \, e^{-i  p\cdot Y } S(Y,K)
\ ,
\label{a31}
\end{eqnarray}
and we finally get the two-particle correlator as
\begin{equation}
C(p,K) = 1\, + \,  \cos{\alpha} \,
\frac{
\left| \int  d^4 X \, e^{i  p\cdot X } S(X,K) \right| ^2 }
{
 \int  d^4 X \, S\Big( X,K+\frac{p}{2} \Big) \,
 \int  d^4 Y \, S\Big( Y,K-\frac{p}{2} \Big)
}
\ ,
\label{a32}
\end{equation}
where the source function $S(X,K)$ is defined in accordance with
Eq.~(\ref{a110}) and all integrations  are taken at emission
times or on the freeze-out hyper-surface.


\end{document}